\documentclass[twocolumn,prl,showpacs,superscriptaddress,preprintnumbers,amsmath,amssymb]{revtex4}

\usepackage[dvips]{graphicx}
\usepackage{dcolumn}
\usepackage{bm}

\begin{document}


\title{Determining the crystal-field ground state in rare earth
Heavy Fermion materials using soft-x-ray absorption spectroscopy}

\author{P. Hansmann}
  \affiliation{II. Physikalisches Institut, Universit{\"a}t zu K{\"o}ln,
   Z{\"u}lpicher Stra{\ss}e 77, D-50937 K{\"o}ln, Germany}
\author{A. Severing}
  \affiliation{II. Physikalisches Institut, Universit{\"a}t zu K{\"o}ln,
   Z{\"u}lpicher Stra{\ss}e 77, D-50937 K{\"o}ln, Germany}
\author{Z. Hu}
  \affiliation{II. Physikalisches Institut, Universit{\"a}t zu K{\"o}ln,
   Z{\"u}lpicher Stra{\ss}e 77, D-50937 K{\"o}ln, Germany}
\author{M. W. Haverkort}
  \affiliation{II. Physikalisches Institut, Universit{\"a}t zu K{\"o}ln,
   Z{\"u}lpicher Stra{\ss}e 77, D-50937 K{\"o}ln, Germany}
\author{C. F. Chang}
  \affiliation{II. Physikalisches Institut, Universit{\"a}t zu K{\"o}ln,
   Z{\"u}lpicher Stra{\ss}e 77, D-50937 K{\"o}ln, Germany}
\author{S. Klein}
  \affiliation{II. Physikalisches Institut, Universit{\"a}t zu K{\"o}ln,
   Z{\"u}lpicher Stra{\ss}e 77, D-50937 K{\"o}ln, Germany}
\author{A.~Tanaka}
  \affiliation{Department of Quantum Matter, ADSM Hiroshima
  University, Higashi-Hiroshima 739-8530, Japan}
\author{H. H. Hsieh}
  \affiliation{Chung Cheng Institute of Technology, National Defense University,
  Taoyuan 335, Taiwan}
\author{H.-J. Lin}
  \affiliation{National Synchrotron Radiation Research Center (NSRRC), 101 Hsin-Ann
  Road, Hsinchu 30077, Taiwan}
\author{C. T. Chen}
  \affiliation{National Synchrotron Radiation Research Center (NSRRC), 101 Hsin-Ann
  Road, Hsinchu 30077, Taiwan}
\author{B. F{\aa}k}
  \affiliation{CEA, D\'epartement de Recherche Fondamentale sur la
  Mati\`ere Condens\'ee, SPSMS, 38054, Grenoble, France}
\author{P. Lejay}
\affiliation{Institut N\'eel, CNRS, BP 166, 38042 Grenoble Cedex
9, France}
\author{L. H. Tjeng}
  \affiliation{II. Physikalisches Institut, Universit{\"a}t zu K{\"o}ln,
   Z{\"u}lpicher Stra{\ss}e 77, D-50937 K{\"o}ln, Germany}

\date{\today}

\begin{abstract}

We infer that soft-x-ray absorption spectroscopy is a versatile
method for the determination of the crystal-field ground state
symmetry of rare earth Heavy Fermion systems, complementing
neutron scattering. Using realistic and universal parameters, we
provide a theoretical mapping between the polarization dependence
of Ce $M_{4,5}$ spectra and the charge distribution of the Ce $4f$
states. The experimental resolution can be orders of magnitude
larger than the $4f$ crystal field splitting itself. To
demonstrate the experimental feasibility of the method, we
investigated CePd$_2$Si$_2$, thereby settling an existing
disagreement about its crystal-field ground state.

\end{abstract}


\maketitle

Heavy Fermion materials are strongly correlated rare earth or
actinide materials where the atomic-like $f$ electrons interact
with conduction electrons giving rise to extraordinary low energy
properties. There has been a revival of interest in many of these
well studied compounds with an intense search for new ones since
the discovery that many of these materials are at the border
between magnetic order and superconductivity when driven through a
quantum critical point by either applying a magnetic field or
pressure, and/or exhibit non-Fermi liquid behavior
\cite{Jaccard1992, Movshovich1996, Mathur1998, Hegger2000,
Movshovich2001, Pagliuso2002, Park2006, Trovarelli2000,
Loehneysen2007}. There are indications that the unconventional
superconductivity in these compounds is triggered by
antiferromagnetic correlations leading to the interesting aspect
of competing magnetic and superconducting order parameters which
may involve a momentum dependent hybridization of $f$ and
conduction electrons \cite{Mena2005, Burch2007, Ghaemi2007}.
Therefore, any microscopic understanding of how the wealth of
properties in Heavy Fermion materials evolves out of the strongly
interacting $f$ ground state requires the knowledge of the spatial
distribution of the $f$ state involved.  It has been pointed out
already early by Zwicknagl, for example, that the Fermi surface of
several systems depends strongly on the crystal-field symmetries
\cite{Zwicknagl1992}. Other effects like the quadrupolar ordering
reflect the charge distribution of the $f$ ground state
\cite{Fazekas, Kiss2003, Kono2004, Kuramoto2005}, but so far
reliable input from experiment is scarce.

The standard experimental technique to determine the crystal-field
energy level scheme is inelastic neutron scattering on
polycrystalline samples. The additional information of quasi- and
inelastic intensities should be sufficient to extract information
about the $f$ spatial distribution as well, but the analysis of
magnetic intensities is often hampered by broadened lines, phonons
in the same energy window as the magnetic excitations or strong
absorption of one of the sample's constituents (e.g. Rh, In, B, or
Sm). Therefore, the wave functions of the crystal-field states are
often determined from a combined analysis of neutron and single
crystal static susceptibility data \cite{Goremychkin1993,
Dijk2000, Abe1998}. However, as pointed out by Witte {\it et al.}
\cite{Witte2007} and Janou$\rm\breve{a}$sova {\it et al.}
\cite{Janou2004}, even this combined method often fails due to the
powder and thermal averaging in either technique. Moreover, in the
presence of magnetic order anisotropic molecular field parameters
have to be introduced in order to fit the static susceptibility,
ending up with too many free parameters for a unique description
of the crystal-field potential, leaving ambiguities and unresolved
debates in the present literature. One then has to resort to time
consuming inelastic polarized neutron scattering experiments with
their demand for large single crystals \cite{Janou2004,
Witte2007}, or, equally time consuming, to elastic scattering
experiments with polarized neutrons to determine the $f$ magnetic
form factor \cite{Boucherle1985}. The latter has the advantage
that it is also applicable for systems in which crystal-field
excitations are not defined \cite{Boucherle2001}. Nevertheless,
neutron based methods have their limitations in case the compound
under study contains more than one type of magnetic ion.

We will show in this report at the example of CePd$_2$Si$_2$ that
polarization dependent soft-x-ray absorption spectroscopy at the
rare earth \emph{M$_{4,5}$} edges (soft-XAS) is a powerful tool to
give undisguised direct information concerning the $4f$ charge
distribution in the ground state. This technique is complementary
to inelastic and elastic neutron scattering, but has the advantage
of having orders of magnitude better
signal-to-noise/background-ratio and of requiring only small
amounts of sample material. It is capable of measuring materials
which are strongly absorbing for neutrons, and it is also element
specific, so that the presence of more than one type of magnetic
ion is no limitation.

\begin{figure}
    \includegraphics[width=0.36\textwidth]{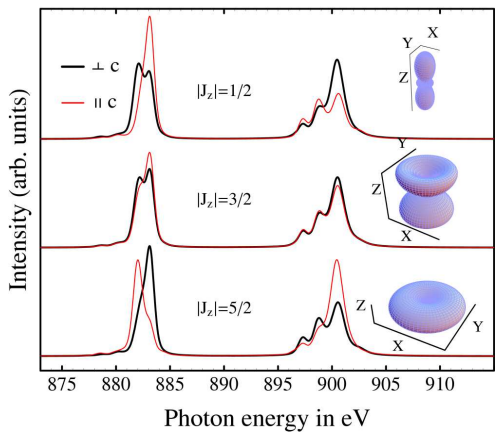}
    \caption{(color online) Calculated $M_{4,5}$ soft-x-ray absorption spectra for pure
     $|$$J_z$$\rangle$ states of Ce$^{3+}$
     for the incoming light polarized $\parallel$ and $\perp$ ${\bf c}$
     (${\bf z}$ $\parallel$ ${\bf c}$ in tetragonal symmetry - see below). On the right
     the spatial distribution of the $4f$ electrons for the respective
     $|$$J_z$$\rangle$ states.}
    \label{fig1}
 \end{figure}

Over the last 20 years soft-XAS has developed into a powerful
technique for studying the element-specific local electronic
structure of $3d$ transition metal and $4f$ rare earth compounds.
The respective $2p$-to-$3d$ ($L_{2,3}$)and $3d$-to-$4f$
($M_{4,5}$) absorption processes can be well understood in terms
of atomic-like transitions into multiplet-split final states
\cite{Thole1985,Tanaka94,deGroot94,Thole97}. Unique is that the
dipole selection rules are very effective in determining which of
those final states can be reached and with what intensity for a
given initial state symmetry. This makes the technique an
extremely sensitive local probe, ideal to study the valence, spin
and orbital state of the ions of interest.

The majority of the soft-XAS studies on rare earth materials were
so far focused on issues concerning their magnetism by measuring
the magnetic circular dichroic effect \cite{Goedkoop1988}.
Surprisingly, only little attention had been paid on orbital
symmetry or crystal-field issues in rare earths: there are a few
soft-XAS reports early on addressing surface and interface
crystal-fields \cite{Sacchi1991,Kappert1993,Castrucci1995}. As far
as bulk materials are concerned, only the giant crystal-field
system CeRh$_3$B$_2$ has been investigated
\cite{Jo1990a,Jo1990b,Yamaguchi1992,Nakai1993,Yamaguchi1995},
thereby giving the impression that it would be difficult to study
experimentally systems with smaller crystal-fields. Actually,
CeRh$_3$B$_2$ is a quite special case since the large size of the
crystal-field splitting is comparable to that of the spin-orbit
interaction, resulting in an intermixing of the $J$=7/2 and
$J$=5/2 orbital states \cite{Givord2004}. Here we follow up on the
theoretical work of Jo {\it et al.} \cite{Jo1990a,Jo1990b} and
show that for the small crystal-fields in rare earth systems a
simple mapping can be found between the ground state orbital
symmetry and the polarization dependence of the spectra, allowing
an accurate and quick quantitative analysis.

The spin orbit interaction in the $f$ shell of rare earth
compounds is much larger than the crystal-field splitting due to
its small spatial distribution. $J$ in the $LS$ coupling is
therefore a good quantum number, and the eigenstates in the
presence of crystal-fields can be expressed in terms of linear
combinations of the corresponding $|$$J_z$$\rangle$ states. In the
case of Ce$^{3+}$ with $f^1$ in tetragonal site symmetry, for
example, the 6-fold degenerate $J$=5/2 state splits into three
doublets: $|$1$\rangle$ = $a$$|$$\pm5/2$$\rangle$ -
$b$$|$$\mp3/2$$\rangle$, $|$2$\rangle$ = $|$$\pm1/2$$\rangle$ and
$|$3$\rangle$ = $b$$|$$\pm5/2$$\rangle$ + $a$$|$$\mp3/2$$\rangle$
with $|\Gamma_7^1$$\rangle$, $|\Gamma_6$$\rangle$, and
$|\Gamma_7^2$$\rangle$ symmetry, respectively. The crucial
question is which one of these states forms the ground state and
what are the $a$ and $b$ coefficients ($a^2$+$b^2$=1), thus
defining the actual charge distribution. In Stevens formalism
\cite{Stevens1951} the tetragonal crystal-field potential of an
$f^1$ is given by the three crystal-field parameters $B_2^0$,
$B_4^0$ and $B_4^4$, and can be expressed alternatively through
the transition energies $\Delta$$E_{12}$ and $\Delta$$E_{13}$, and
the coefficient $a$.

To study the sensitivity of the Ce \emph{M$_{4,5}$} spectra to the
$4f$ charge distribution of the ground state, we performed atomic
calculations which include the full multiplet theory
\cite{Thole1985,Tanaka94,deGroot94,Thole97,Goedkoop1988} using the
XTLS 8.3 programm \cite{Tanaka94}. It accounts for the
intra-atomic $4f$-$4f$ and $3d$-$4f$ Coulomb and exchange
interactions, the $3d$ and $4f$ spin-orbit couplings, and the
local crystal-field parameters. All atomic parameters were given
by the Hartree-Fock values, with a reduction of about 35\% for the
$4f$-$4f$ parameters and 25\% for the $3d$-$4f$ to reproduce best
the experimental isotropic spectra of Ce materials with low Kondo
temperatures, as to account for configuration interaction effects
not included in the Hartree-Fock scheme \cite{Tanaka94}.

We start with calculating the \emph{M$_{4,5}$} spectra for the
pure $|J_z\rangle$ states of the Ce$^{3+}$ with $4f^1$ and $J$=5/2
for two polarizations of the electric field vector $\vec{E}$,
namely $\vec{E} \parallel {\bf c}$ and $\vec{E} \perp {\bf c}$,
where the tetragonal ${\bf c}$-axis is aligned along the ${\bf z}$
direction. Fig. 1 shows the results of these calculations. The
spectra are dominated by the Ce $3d$ core-hole spin-orbit coupling
which splits the spectrum roughly in two parts, namely the $M_{5}$
($h\nu \approx 877-887$ eV) and $M_{4}$ ($h\nu \approx 895-905$
eV) white lines regions. Crucial is that there is a strong
polarization dependence which is characteristic for each of the
states. The existence of such a polarization dependence is
intuitively clear when looking at the spatial distributions of the
$f$ electrons of the respective $|$$J_z$$\rangle$ states as shown
also in Fig.1: the distributions differ remarkably between the
different $|$$J_z$$\rangle$ states.

\begin{figure}
    \includegraphics[width=0.36\textwidth]{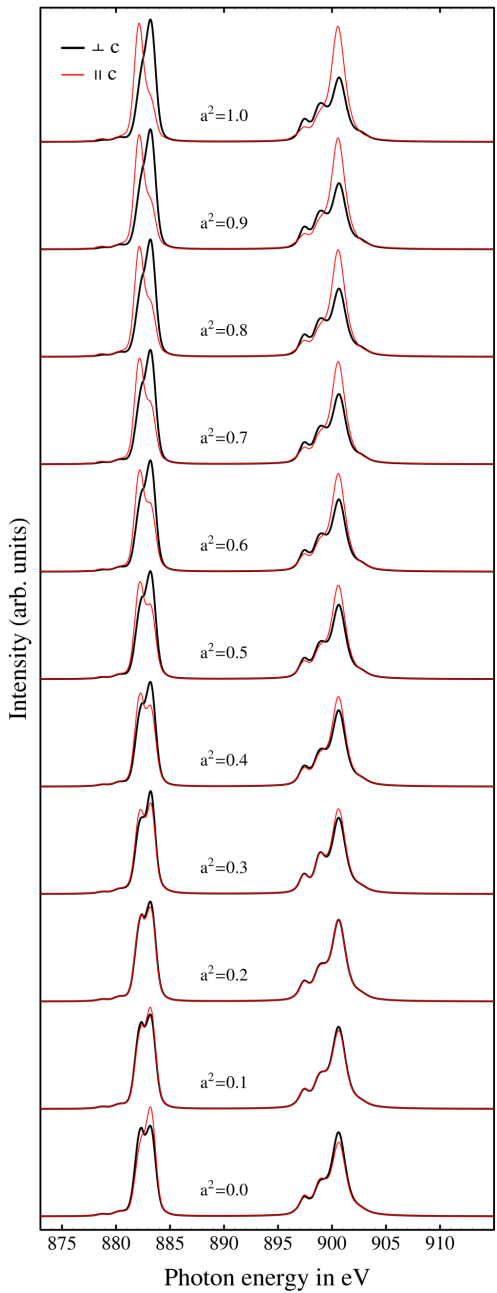}
    \caption{(color online) Top: Calculated $M_{4,5}$ soft-x-ray absorption spectra for the mixed
     states $a$$|$$\pm5/2$$\rangle$ - $b$$|$$\mp3/2$$\rangle$ for mixing
     factors $a^2$ from 0 to 1 for light polarized $\parallel$ and
     $\perp$ ${\bf c}$.}
   \label{fig2}
\end{figure}

The results in Fig.1 are a nice demonstration of the so-called
initial state symmetry effect, an important concept for the
determination of the ground state orbital symmetry in real
materials: the splitting between the various crystal-field states
can be small, e.g. smaller than the broadening of the spectra due
to lifetime, phonon or experimental resolution, and yet the
polarization dependence can be very large. Important is that the
temperature is sufficiently low so that primarily only the lowest
state is populated and contributes to the signal. In fact, we can
make use of the fact that the crystal-field splittings in rare
earth materials are typically much smaller than the $\approx$400
meV spectral broadening. When calculating the polarization
dependence of the states $|$1$\rangle$ or $|$3$\rangle$, we find
they can be well approximated by an incoherent sum of the
$|$3/2$\rangle$ and $|$5/2$\rangle$ spectra, weighted with the
proper $a^2$ and $b^2$ parameters. The results are shown in Fig.
2, where we let $a^2$ run from 0 to 1, with $a^2$=0 corresponding
to a pure $|$3/2$\rangle$ and $a^2$=1 to a pure $|$5/2$\rangle$
state. Fig. 2 in effect can serve as a map to find the coefficient
$a^2$ directly from an experiment. It should be generally
applicable for tetragonal Ce materials without the need to know
the crystal-field splittings themselves as long as the total
crystal-field splitting is small in comparison to the spin orbit
splitting of 230 meV. Yet, having knowledge of the crystal-field
energies e.g. from neutron scattering, this XAS technique will
also enable us to give a clear and unique set of crystal-field
parameters, independent of molecular fields. Important is, of
course, that this atomic approach is only accurate for materials
in which the Kondo temperature is sufficiently smaller than the
splitting between the lowest and first excited crystal-field
state, otherwise the analysis needs to be extended by using the
Anderson impurity model.

\begin{figure}
    \includegraphics[width=0.36\textwidth]{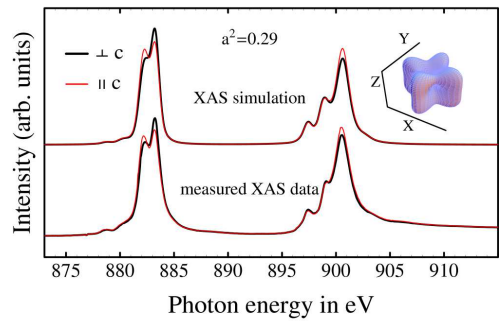}
     \caption{(color online) Bottom: experimental $M_{4,5}$ soft-x-ray absorption
     spectra of CePd$_2$Si$_2$ at T = 50 K for light polarized $\parallel$ and
$\perp$ ${\bf c}$.
     Top: simulated $M_{4,5}$ spectra with $a^2$ =0.29.}
    \label{fig3}
 \end{figure}

We have chosen to measure the linear polarized soft-XAS spectra of
CePd$_2$Si$_2$ in order to proof the feasibility of the technique.
Several groups measured the crystal-fields scheme of
CePd$_2$Si$_2$ with neutron scattering \cite{Dijk2000, Abe1998,
Severing89b, Steeman1990} and agree in the energy splitting but
end up with considerably different ground state wave functions.
The authors of \cite{Dijk2000} and \cite{Abe1998} both took the
energy splitting from neutron data and fitted the single crystal
static susceptibility in order to obtain the ground state wave
function. Both groups obtained decent fits although different
ground state coefficients were assumed. This is due to the fact
that anisotropic molecular field parameters were introduced in
order to take into account the antiferromagnetic order at 10 K,
ending up with too many free parameters for a unique fit. The
CePd$_2$Si$_2$ single crystal used was grown by the Czochralski
method as described in \cite{Dijk2000}. It orders at $T_N$
$\approx$ 10 K. The XAS measurements were performed at the Dragon
beamline of the NSRRC in Taiwan. The spectra were recorded using
the total electron yield method in an ultra-high-vacuum chamber
with a base pressure of 2x10$^{-10}$ mbar. Clean sample areas were
obtained by cleaving the crystals \textit{in-situ}. The photon
energy resolution at the Ce $M_{4,5}$ edges ($h\nu \approx
875-910$ eV) was set at 0.4 eV, and the degree of linear
polarization was $\approx 98 \%$. The CePd$_2$Si$_2$ single
crystal was mounted with the ${\bf c}$-axis perpendicular to the
Poynting vector of the light. By rotating the sample around this
Poynting vector, the polarization of the electric field vector can
be varied continuously from $\vec{E}
\parallel {\bf c}$ to $\vec{E} \perp {\bf c}$. This measurement geometry
allows for an optical path of the incoming beam which is
independent of the polarization, guaranteeing a reliable
comparison of the spectral line shapes as a function of
polarization.

\begin{figure}
    \includegraphics[width=0.36\textwidth]{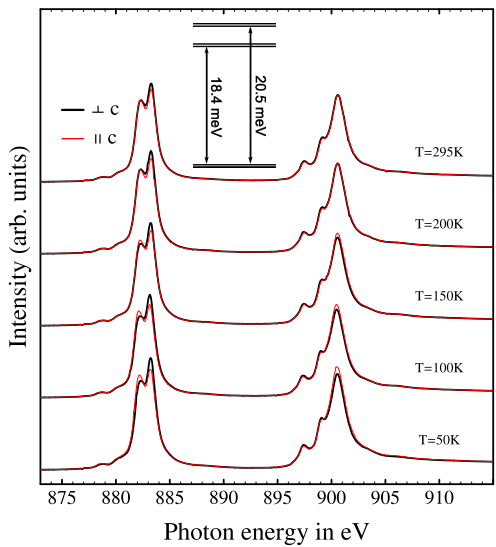}
     \caption{(color online) Temperature dependence of the experimental
     $M_{4,5}$ soft-x-ray absorption spectra of CePd$_2$Si$_2$.
     The inset gives the crystal-field transition
     energies as in \cite{Severing89b}}
    \label{fig4}
 \end{figure}

The bottom curves of Fig. 3 show the measured soft-XAS spectra of
CePd$_2$Si$_2$ at T = 50 K for the light polarized parallel and
perpendicular to the ${\bf c}$-axis. The spectral features are
well resolved and their sharpness can be taken as a confirmation
for the atomic character of the Ce states. Important is that there
is a polarization dependence and that this can be clearly seen due
to the excellent signal-to-noise/background ratio. Comparing these
experimental spectra to the simulations in Figs. 1 and 2, one can
immediately deduce that the ground state must be state
$|$1$\rangle$ with $a^2$ close to 0.3. Indeed, a more detailed
simulation analysis yields the best fit for $a^2$=0.29. We know
from neutron data \cite{Severing89b, Steeman1990} that the first
excited crystal-field level is above 18 meV, i.e. well above 200
K, so that at T = 50 K only the ground state contributes to the
spectra. When simulating the data the Boltzmann factor was
nevertheless taken into account. The value found for $a^2$
corresponds to $a$=0.54 and $b$=0.84, so that we essentially
confirm the ground state of CePd$_2$Si$_2$ as proposed by
\cite{Dijk2000}: $|$1$\rangle$ = 0.55 $|$$\pm5/2$$\rangle$ - 0.84
$|$$\pm3/2$$\rangle$. The corresponding spatial distribution of
the $4f$ electron in this ground state is shown in the inset of
Fig. 3.

To confirm the crystal-field nature of the polarization effect, we
have also carried out measurements at elevated temperatures. The
polarization dependence in the spectra diminishes as the
temperature rises as is shown in Fig. 4 for 50, 100, 150, 200 and
295 K. This is in accordance with the increased population of
higher crystal-field levels, resulting eventually in an isotropic
spectrum at sufficiently high temperatures. These elevated
temperature spectra can be well described using the ground state
symmetry and coefficient from the 50 K spectrum and the
crystal-field energies from neutron data \cite{Severing89b},
depicted in the inset of Fig. 4. One only has to take into account
the Boltzmann factor for each respective temperature.

In conclusion we have shown that the Ce $M_{4,5}$ spectra have a
polarization dependence which is characteristic for the charge
distribution of the Ce $4f$ states. We have provided a set of
reference spectra (Figs. 1 and 2) from which one can find the
ground state parameters directly from the experiment, without the
need to know the crystal-field energies themselves as long as they
are small. Using this method we have investigated the local
electronic structure CePd$_2$Si$_2$ and settled the debate
concerning the ground state wave function of this compound,
demonstrating that soft-x-ray absorption spectroscopy is a
powerful method complementing neutron scattering.

We would like to thank Lucie Hamdan for her skillful assistance in
preparing the experiments, and John Mydosh and Daniel Khomskii for
valuable discussions.


\end{document}